# Gender-Specific Patterns in the Artificial Intelligence Scientific Ecosystem

Anahita Hajibabaei[1], Andrea Schiffauerova[1], and Ashkan Ebadi[1,2,*]

[1] Concordia University, Concordia Institute for Information Systems Engineering, Montreal, QC H3G 2W1 Canada
[2] National Research Council Canada, Montreal, QC H3T 1J4, Canada
[*] ashkan.ebadi@nrc-cnrc.gc.ca

**Abstract**

Gender disparity in science is one of the most focused debating points among authorities and the scientific community. Over the last few decades, numerous initiatives have endeavored to accelerate gender equity in academia and research society. However, despite the ongoing efforts, gaps persist across the world, and more measures need to be taken. Using social network analysis, natural language processing, and machine learning, in this study, we comprehensively analyzed gender-specific patterns in the highly interdisciplinary and evolving field of artificial intelligence for the period of 2000-2019. Our findings suggest an overall increasing rate of mixed-gender collaborations. From the observed gender-specific collaborative patterns, the existence of disciplinary homophily at both dyadic and team levels is confirmed. However, a higher preference was observed for female researchers to form homophilous collaborative links. Our core-periphery analysis indicated a significant positive association between having diverse collaboration and scientific performance and experience. We found evidence in support of expecting the rise of new female superstar researchers in the artificial intelligence field.

## 1. Introduction

During the past decades, gender disparity in science has become one of the most focused debating points among researchers (Shen, 2013). Despite concerted efforts to achieve gender equality (UNESCO, 2018), conclusive evidence has revealed that pervasive gender imbalances still exist in academic and scientific practice (Nelson & Rogers, 2003; Shaw & Stanton, 2012; West et al., 2013). Although the number of female and male students pursuing bachelor's and master's degrees is almost equal, the women's under-representation starts at the Ph.D. level, and female attrition can be observed across higher levels of the academic hierarchy - a phenomenon dubbed the *"leaky pipeline"* (Huyer, 2018). Abundant studies have also shown persistent gender bias in many aspects such as salaries (Shen, 2013), hiring and promoting (Moss-Racusin et al., 2012; Nelson & Rogers, 2003; Nielsen, 2016), grant funding (Witteman et al., 2019), scholarly authorship (West et al., 2013), scientific impact (Larivière, Ni, et al., 2013), peer reviews (Murray et al., 2019), and collaboration (Uhly et al., 2015; Zeng et al., 2016).

There are many underlying reasons for the aforementioned imbalances. Family and caring responsibilities have been noted as the main impediment to female academic trajectories (Stack, 2004). While men follow a linear career path, women are more prone to career interruption or relinquishment due to maternity and family issues (Ramos et al., 2015), which could also lead to a gender gap in scientific performance (Kyvik & Teigen, 1996). Another viewpoint focusing on personal preferences or innate individual abilities (M. T. Wang & Degol, 2017) may explain the two genders' academic career dichotomy. Research has also reported gender stereotypes (Ceci et al., 2014), inadequate female role models (Nelson & Rogers, 2003), and the degree of specialization (Leahey, 2006) as potential explanations for the gender bias; however, the extent to which factors matter for this issue has remained elusive among researchers (Shaw & Stanton, 2012).

Over the last few decades, numerous initiatives have endeavored to accelerate gender equity in academia. Several countries enacted effective policies such as introducing flexible and family-oriented programs in the workplace, e.g. in the European Union (EU) countries, India, and Japan (Bonetta & Clayton, 2008; Jurviste et al., 2016), providing financial incentives for appointment and retention of female scientists (Pearson et al., 2015), and promoting gender-balanced organizational structures (Kamraro, 2014). Canada has not been an exception, generous parental leave programs were introduced, including extended parental leave for graduate students and parental sharing benefits, thereby fostering greater equality in family responsibilities and facilitating career recruitment and progression for females (Government of Canada, 2019). Additionally, the Canadian government has recently initiated an internationally recognized program called *"Athena Scientific Women's Academic Network"* (Government of Canada, 2018), aiming to enhance women's career trajectories and adopt an attitude towards gender parity (Ovseiko et al., 2017). Despite these ongoing efforts and policies, gaps persist across the world, and more measures need to be taken on this matter (UNESCO, 2018).

There is a significant body of evidence aiming to assess various aspects of gender differences in academic performance. It has been argued that women are generally less productive than men (Ghiasi et al., 2015; Huang et al., 2020; Larivière, Ni, et al., 2013; Xie & Shauman, 1998), in terms of the number of publications, and this gap varies across research disciplines (Duch et al., 2012; Elsevier, 2017). Myriad studies have been devoted to analyzing and mapping scientific performance and gender patterns through the number of citations. Nonetheless, the research on this area has produced inconclusive and diverse results. For example, while some studies found a citation bias favoring female authors in specific fields (Borrego et al., 2010; Long, 1992; Van Arensbergen et al., 2012), others found no differences for women or men (Bordons et al., 2003; Cole & Zuckerman, 1984; Mauleón et al., 2008; Tower et al., 2007). However, several studies concluded that female authors, on average, have a lower impact compared to male authors (Hunter & Leahey, 2010; Larivière, Diepeveen, et al., 2013; Larivière et al., 2011; Sugimoto, 2013). Recent research on longitudinal gender disparities in scientific performance has revealed that although women's participation has increased in academia over the past 60 years, gender inequality in publications and citation rates has been growing in favor of men. It is also argued that this gap persists among disciplines and almost all countries and is exacerbated when it comes to highly productive researchers (Huang et al., 2020).

It is widely acknowledged that collaboration has become a hallmark of scientific activities (Cummings et al., 2005; Wood & Gray, 1991). Collaborative research is not only crucial to address complex scientific issues (Bennett & Gadlin, 2012) but also may affect research performance (Ebadi & Schiffauerova, 2016; Uhly et al., 2015), professional recognition (Fox & Mohapatra, 2007), and funding opportunities (Ebadi & Schiffauerova, 2015b; Kwiek, 2020). Owing to the importance of scientific collaboration, existing literature has examined gender-specific patterns in academic collaborations and reported gender homophily effect, i.e. if same-gender authors tend to publish together (Holman & Morandin, 2019; Jadidi et al., 2018; Lee et al., 2019), and female propensity to domestic and intramural collaboration (Abramo et al., 2013; Bozeman & Corley, 2004; Larivière et al., 2011; Larivière, Ni, et al., 2013). Furthermore, whilst women opt to form smaller and short-lived ties, men have more extensive and loyal collaboration networks (Jadidi et al., 2018). Presumably, several causes could be attributed to these differences like family obligations, prejudices against women (Hogan et al., 2010), and gender inequality in research funding (Larivière et al., 2011; Stack, 2004).

Although prior studies investigated sex differences in research collaboration and productivity, to the best of our knowledge, no study hitherto addresses gender differences in the propensity to collaborate with respect to the similarity or dissimilarity of researchers' academic backgrounds. In this work, we aim to scrutinize the presence of gender-specific patterns in the co-authorship networks in the field of artificial intelligence (AI), an interdisciplinary field with a significant gender gap (World Economic Forum, 2018). To this end, we utilized machine learning, natural language processing, and social network analysis to address the following research questions: (Q1) Do female and male scientists prefer gaining a wide range of experience in different AI subject areas, or they tend to concentrate more on limited fields?; (Q2) Does disciplinary homophily or diversity influence female and male scientists' preferences to form interdisciplinary research collaboration?; (Q3) What is the association between researchers' position in the network and their scientific performance and is it different for female and male researchers? The remainder of the paper is organized as follows: Section 2 describes data and methodology; Section 3 presents empirical results; Section 4 discusses the findings; Section 5 concludes the article; and Section 6 represents the limitations of this study and draws some directions for future research.

## 2. Data and Methods

### 2.1 Data

Data collection and preparation involved several steps. First, the bibliographic data including but not limited to title, abstract, keywords, date of publication, author list, etc. were retrieved from Elsevier's Scopus, filtering in research articles, conference papers, book chapters, and books published from 2000 to 2019. We only included publications for which both title and abstract were available. We used the ("artificial intelligence" OR "machine learning" OR "deep learning") search query to extract AI-related publications where at least one of the mentioned phrases appeared in the title/abstract of the publication or in the keywords section. Using machine learning techniques and natural language processing, we then coded and applied an automatic gender assignment model trained on a large labeled dataset of names to infer researchers' gender from a set of primary features such as their full name, affiliation, and country of origin. We classified authors' genders into female (F), male (M), and unisex/unknown (U). We labeled authors' genders as unisex/unknown (U) when our method could not detect the gender and excluded these authors in further analysis. We carefully validated the accuracy of our automatic gender identification algorithm through manual verification on a random sample of 1000 scientists. The algorithm identifies the gender of the researchers with 96% accuracy (94% for females, 98% for males). The final dataset contains 39,679 publications excluding unknown/unisex only articles as well as those without abstract, and 114,371 authors (30,448 females and 83,923 males).

We then used social network analysis (SNA) to build the co-authorship network of AI researchers. For this purpose, we first constructed a bipartite co-authorship network, with authors and articles as the two types of nodes, and edges indicating the publication's authorship (De Nooy et al., 2005). In order to track gender-specific collaborative patterns at the author level, we transformed the bipartite network into a monopartite network wherein nodes represent authors characterized by gender, and two authors were connected if they had a joint publication.

We adopted a topic modeling approach so as to infer the authors' scientific research themes and domains of interest and identify the degree of interdisciplinarity at the author level. The resulted topic vectors from the topic model would represent disciplinary profiles of the scientists in our target dataset, i.e. the AI-related research publications. Whereas journal disciplinary classification and departmental affiliations may be used to recognize the disciplinary profile (Schummer, 2004; J. Wang et al., 2015), they cannot accurately represent the authors' academic backgrounds. For instance, computer scientists who apply their knowledge in various fields of studies like mathematics, physics, medicine, etc. cannot be fully and precisely represented by a merely single computer science discipline. On the other hand, academic publications could properly manifest the interdisciplinarity of individual researchers (Porter et al., 2007). Taken together, we decided to employ the Latent Dirichlet Allocation (LDA) topic modeling technique, developed by Blei et al. (2003), to derive the researchers' disciplinary profiles from their past publications. We first merged the titles and abstracts of the articles. The rationale for this integration is that titles cover specific keywords to index papers, while abstracts contain much more succinct information and represent the research's main idea (Ebadi et al., 2020). To apply the LDA model, several preprocessing steps were carried out on the corpus including the transformation of the textual data to lowercase, tokenization, removal of non-alphabetic characters, removal of the words with length less than three characters, and eliminating custom stop words. Next, we created a document-term matrix to prepare the data for topic modeling. We built several LDA models with different numbers of topics and then calculated and analyzed several evaluation metrics, such as perplexity and log-likelihood, to assess the quality of the models (Griffiths & Steyvers, 2004). Besides quantitative metrics, three domain experts evaluated the quality of the model by analyzing top topic keywords and document-topic distributions and concluded that the optimal number of topics for our research objective is 8. As a result of the LDA model, each publication can be associated with more than one topic with a certain probability. Using the document-topic probability matrix generated by the LDA algorithm, we identified the authors' research fields based on average topic distribution over their past publications. Each author was then represented as a topic distribution vector consisting of 8 disciplines, where each component corresponded to the average topic distribution of the author's past articles under the given field. Moreover, we assumed the topic with the highest probability in the topic distribution vector for each author defines the author's primary discipline. For example, assume that author $i$ has two publications $P_1$ and $P_2$ with document-topic probability vectors, including 8 research themes, as follows: $P_1$(0.5, 0, 0.1, 0, 0, 0.4, 0, 0) and $P_2$(0.6, 0, 0.2, 0.2, 0, 0, 0, 0), where elements in the vector represent the probability that the publication contains the research theme. We calculated the average topic distribution over author $i$'s past publications ($P_1$ and $P_2$) and represented author $i$'s disciplinary profile by a topic distribution vector $\vec{x}_i$(0.55, 0, 0.15, 0.1, 0, 0.2, 0, 0), in which each component shows the probability that author $i$ published under the given discipline.

## 2.2 Methods

### 2.2.1 Descriptive Indicators

To examine the impact of network properties on scientific performance and experience, we calculated and utilized several initial descriptive indicators such as the total number of publications published by each author as a proxy for scientific output and the total number of citations received by each article to measure research impact. It is worth noting that publications published in various fields or years accrue citations differently (Waltman, 2016), wherefore, to correct the effects of field and year of publication, we also calculated field normalized citation counts, which is defined by dividing the total citation counts received by each article by the average citation counts for all articles published in the same year and the same field (Paul-Hus et al., 2015). In addition, career length was used as a proxy for academic experience, defined as the difference between the authors' first and last publications' dates in the dataset.

*2.2.2 Structural Metrics*

Having calculated the set of descriptive indicators, we largely adopted the framework proposed by Feng and Kirkley (2020) to calculate the structural metrics, and did some modifications to satisfy our research objectives. They presented several metrics to capture the degree of disciplinary diversity among researchers at the uni-author, bi-author, and team levels. In this study, we extended their work by taking the gender aspect into account, aiming to assess the existence of disciplinary diversity or homophily among female and male AI researchers in a larger dataset and investigate gender-specific patterns in interdisciplinary collaborations. Moreover, we improved their approach by applying the probabilistic topic modeling technique (LDA) to better capture the research topics and expertise of scientists based on their research output. Following (Feng & Kirkley, 2020), we utilized network structure analysis to categorize researchers based on their collaboration patterns and explore how different collaborative behavior could affect research performance and experience. We introduce each metric in detail in the rest of this section.

*Uni-Author Level Disciplinary Diversity*

We applied Shannon's entropy, i.e. one of the common indices to assess diversity (Aydinoglu et al., 2016; Feng & Kirkley, 2020; Gray, 2011) to measure male and female scientists' disciplinary diversity. Shannon's entropy index is used to measure the dispersion among various fields of the individual author's research background. The degree of interdisciplinarity at the single author level is then identified by the diversity of fields represented in his/her publications, which is defined by the topic distribution vector $\vec{x}_i$. Hence, the entropy of each author's publication history is computed as follows:

$$H_i = -\frac{1}{log(n_d^{(i)})} \sum_{d=1}^{N_d} \vec{x}_{id}\ log(\vec{x}_{id}) \qquad (1)$$

where $n_d^{(i)}$ is the number of unique fields in which researcher $i$ published, and $\vec{x}_{id}$ represents the author $i$'s disciplinary profile based on the average topic distribution of all past publications classified under discipline $d$ (the $d^{th}$ entry of topic distribution vector $\vec{x}_i$). We divided the Shannon entropy metric by the maximum possible diversiy index, i.e. , to normalize the equation (1) between 0 (no diversity) and 1 (highest diversity) and compensate for the variation in the number of disciplines in which researchers contribute to (Shannon, 1948). One may note that authors who contributed to only one field ($n_d = 1$) were excluded from the calculation. Accordingly, an author with a high value of entropy would have more equal publication distribution across different disciplines as well as a higher degree of interdisciplinarity and vice versa. To summarize, the uni-author level diversity metric represents the degree of concentration of researchers in different research themes based on the content of their publications.

*Bi-Author Level Disciplinary Diversity*

To explore whether disciplinary homophily or diversity can be observed as a determining factor in forming a collaborative network in interdisciplinary research, we employed a pairwise similarity measure. The relationship between a pair of researchers can be indicated by edges in the scientific collaboration network, and it could be between authors from similar or diverse academic backgrounds; thus, the more disparate the co-authors' disciplinary profiles, the more heterogeneous their relationships are. To represent dyadic interactions between two genders, we classified edges of the co-authorship network into female-female (FF), female-male (FM), and male-male (MM) collaboration. Then as suggested by Feng and Kirkley (2020), using researchers' topic distribution vectors ($\vec{x}$), we applied the cosine similarity to measure the degree of similarity between a pair of authors in terms of shared interdisciplinary profile as defined in equation (2). Cosine similarity calculates the cosine of two non-zero vectors of *n* dimensions (Han et al., 2011). In this case, the two vectors contain the disciplinary profiles of authors.

$$S_{ij} = \frac{\vec{x}_i \cdot \vec{x}_j}{\|\vec{x}_i\|\|\vec{x}_j\|} \qquad (2)$$

In the equation, $\|\vec{x}\|$ is the norm of the vector $\vec{x}$, and $\vec{x}_i \cdot \vec{x}_j$ denotes the inner product of $\vec{x}_i$ and $\vec{x}_j$. The cosine similarity measure ($S_{ij}$) indicates the similarity between the research profile of author $i$ and $j$, and varies between 0 (authors having collaboration only with those from disparate research backgrounds) and 1 (authors having collaboration only with those with the same research backgrounds).

*Team-Level Disciplinary Diversity*

Team-level disciplinary diversity can be characterized by either the variety of team members' disciplines, i.e. interpersonal diversity, or the degree of individuals' disciplinary diversity, i.e. intrapersonal diversity (Bunderson & Sutcliffe, 2002). In different terms, a diverse team can be composed of researchers from various disciplines or different interdisciplinary individuals (Wagner et al., 2011), or a combination of both of them. Given such a definition, to investigate gender differences in the tendency to collaborate in interdisciplinary research, we followed a similar approach in (Feng & Kirkley, 2020) and considered both primary disciplines and authors' disciplinary profiles as proxies for interdisciplinarity. Additionally, to compare between two gender groups, we considered the authors who publish the same paper as a team and partitioned teams into three categories, labeled as female-only, male-only, and mixed-gender. Then for each team, within-group entropy is defined based on co-authors' primary disciplines to measure team-level interdisciplinarity. The within-group entropy of publication $p$ is defined as:

$$\tilde{H}_P = -\frac{1}{log(min\{|p|, N_d\})} \sum_{d=1}^{N_d} f_{pd}\, log(f_{pd}) \qquad (3)$$

where $f_{pd}$ is the fraction of authors with primary discipline $d$ (the discipline with the highest probability in the topic distribution vector $\vec{x}$) in paper $p$, $N_d$ denotes the number of unique disciplines within each team, and $|p|$ is the total number of authors in a given paper. As a tight upper bound on the entropy is restricted either by $N_d$ or $|p|$, a normalization factor of $(log(min\{|p|, N_d\}))^{-1}$ was introduced in Equation (3).

Similarly to equation ( *1* ), a high value of within-group entropy suggests a high degree of disciplinary diversity at the team level. Furthermore, we applied average within-group cosine similarity to measure the level of team interdisciplinarity based on team members' disciplinary profiles, i.e. topic distribution vectors. The average within-group cosine similarity for each publication is expressed as:

$$\tilde{S}_p = \frac{2}{|p|(|p|-1)} \sum_{(i,j) \in p} S_{ij} \qquad (4)$$

where $|p|$ is the total number of authors in a given paper, $S_{ij}$ is the cosine similarity between author $i$ and $j$ vectors, and the summation is taken over all pairs of authors in the paper $p$. Using $2\big(|p|(|p|-1)\big)^{-1}$, we normalized the $\tilde{S}_p$ values between 0 and 1, 1 indicates the maximum team disciplinary homophily. In summary, Equations (3) and (4) help us to explore interdisciplinarity at the team level from two perspectives: (1) based on the diversity of disciplines across team members, utilizing the within-group entropy, and (2) based on the disciplinary diversity of individuals within the team, using the average within-group cosine similarity.

*Core-Periphery Researcher Decomposition*

Researchers' position in scientific collaboration networks could affect their performance (Ebadi & Schiffauerova, 2015b, 2016). Therefore, we investigated the relationship between researchers' brokerage role and their academic performance and examined how this relationship could differ between men and women. We used betweenness centrality as a proxy for researchers' influence, importance, and their brokerage role in the network. This measure can identify researchers who act as intermediaries between different groups of researchers, called gatekeepers. Gatekeepers can play a crucial role in the network by connecting different clusters and transferring and/or controlling the flow of information/knowledge between communities (Ebadi & Schiffauerova, 2015a). Betweenness centrality of author $i$ is defined based on the number of times that author $i$ connects two other authors via the shortest path passing through author $i$ (Borgatti, 2005). Hence, the betweenness centrality of the given node ($bc_i$) is defined as:

$$bc_i = \sum_{i \neq j \neq k} \frac{\sigma_{jk}(i)}{\sigma_{jk}} \qquad (5)$$

In Equation (5), $\sigma_{jk}$ denotes the total number of the shortest paths between node $j$ and $k$, and $\sigma_{jk}(i)$ is the number of those paths containing node $i$. The betweenness centrality value ranges from 0 to 1, and the higher the value is, the more influence the node has. We performed core-periphery analysis as proposed by Feng and Kirkley (2020) to distinguish researchers based on their influence within the network. We utilized different network measure from theirs and calculated betweenness centrality to determine the most central and influential female and male scientists and categorized nodes into two roles: (1) The upper 5% of authors in terms of betweenness value are associated with the "*core*" role in the network, and (2) the remaining 95% as the "*periphery*" role. This helped us to identify core researchers who act as a bridge between different communities and funnel the information in the network. Due to their roles, they may benefit from having more opportunities to collaborate more and produce higher impact research (Abbasi et al., 2012; Yan & Ding, 2009). Figure 1 illustrates the conceptual flow of the study.

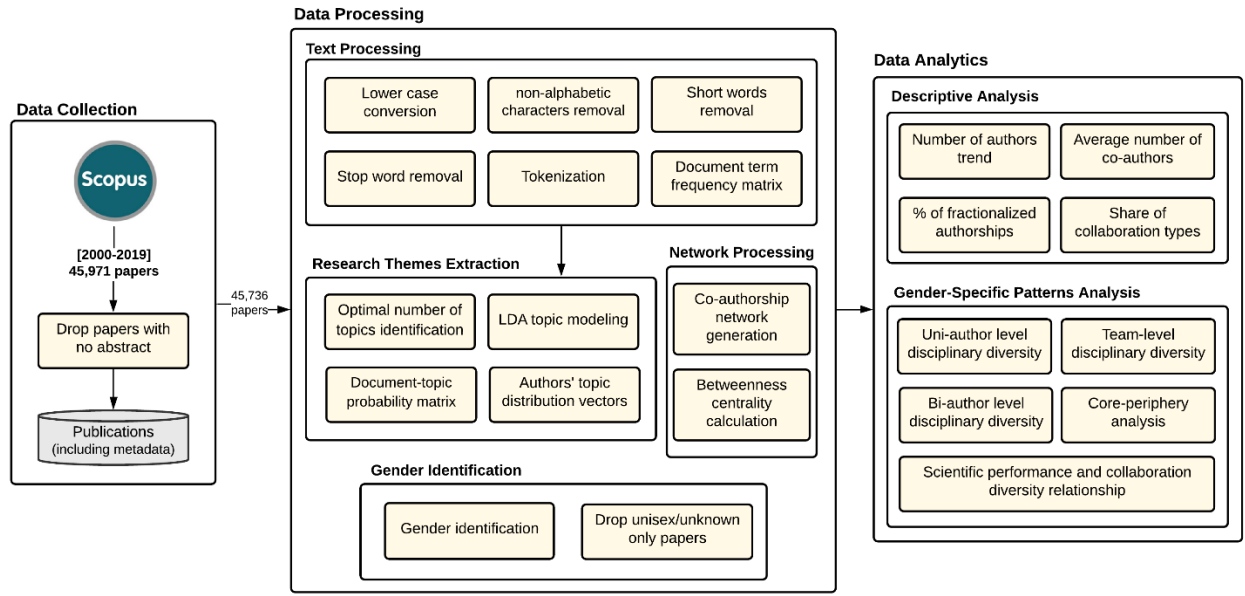

**Figure 1.** The conceptual flow. The pipeline contains three main components, i.e. data collection, data processing, and data analytics. In the data collection component, publications, and their metadata such as title, abstract, keywords, date of publication, author list, etc. are collected from 2000 through 2019. In the data processing component, first, the text data (publications' title + abstract) is processed and used to extract authors' research themes represented by topic distribution vectors. Next, the gender of the authors is identified, their co-authorship network is generated, and their betweenness centrality measure is calculated. The target data is finally passed to the data analytics component to analyze gender-specific patterns within the co-authorship network.

## 3. Results

### 3.1 Descriptive Analyses

The constructed co-authorship network contained 83,923 men and 30,448 women who contribute to 39,679 publications throughout 2000-2019. As seen in Figure 2-a, the number of female and male researchers has been increasing steadily over the years, following an almost similar pattern. Despite the growth, female researchers are still fewer, overall making up only ~27% of the AI research community based on the publications' authorship. We also compared publication outputs of the two genders using fractional authorship counts, where each author is given *1/n* credit for the authorship of a given article where *n* is the number of co-authors in the article (Figure 2-b). Although female authors' contribution to published articles is slightly increasing over time, they accounted for less than 30% of fractionalized authorships during the whole period. This partially implies that the gender gap persists in the AI research community.

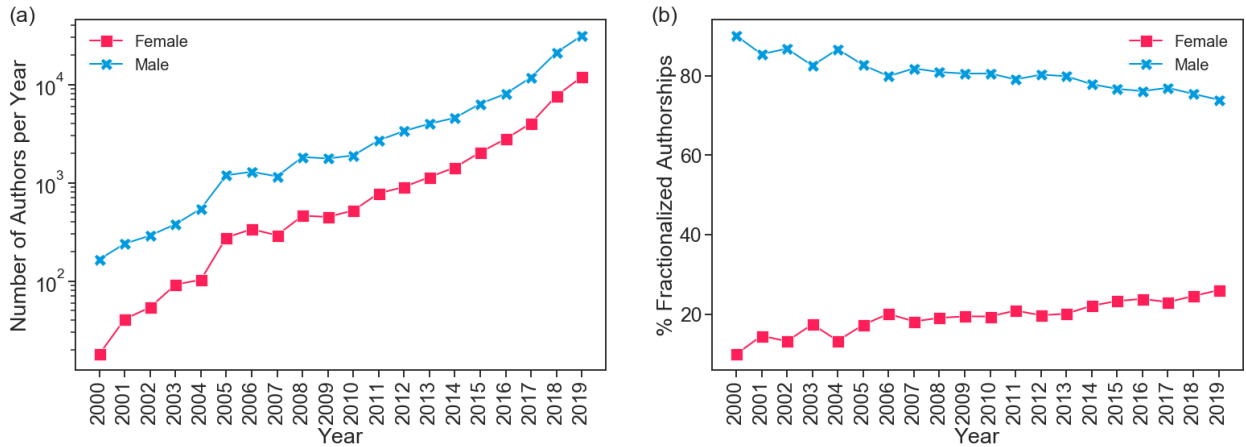

**Figure 2. a)** The total number of female and male authors per year, and **b)** the percentage of fractionalized authorships for two genders.

Figure 3-a shows the average number of distinct co-authors per paper for each gender (left y-axis) as well as the average total team size trend defined by the number of authors per paper (right y-axis). According to Figure 3-a, females have more distinct co-authors on average than their male counterparts, which is consistent with previous studies reporting that women are more collaborative than men (Abramo et al., 2013; Ghiasi et al., 2018). Figure 3-b shows the trends of different collaboration types. From the figure, it can be inferred that the share of male-male (MM) collaboration has followed a downward trend, whereas the shares of female-female (FF) and mixed-gender (FM) collaborations have been increasing over the years. This implies a change in collaboration preferences among the two gender groups and the increasing role and importance of female researchers in the AI community.

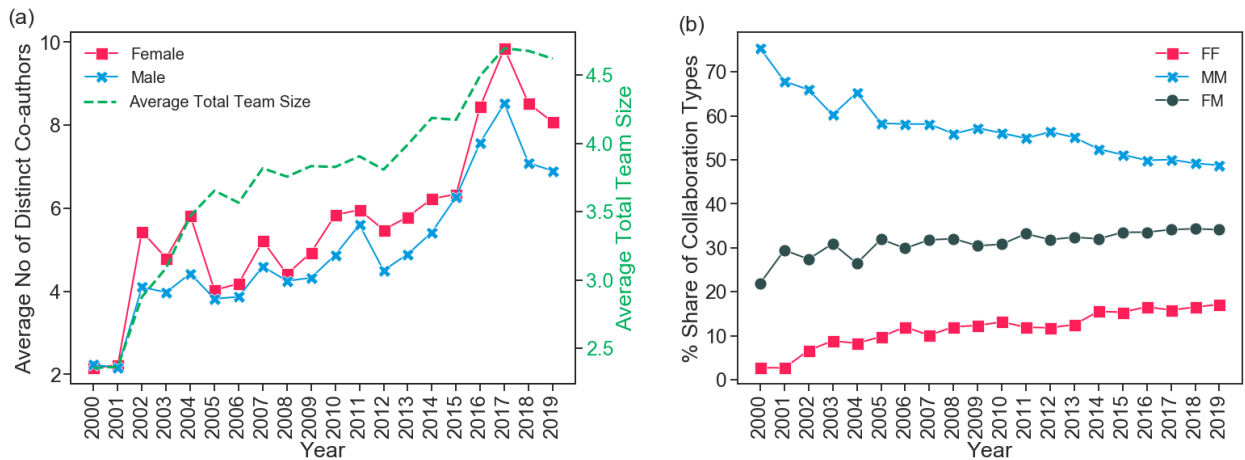

**Figure 3. a)** The average number of distinct co-authors per paper for each gender (left y-axis) and the average number of authors per paper, i.e. average total team size (right y-axis, which is shown by green dashed line), and **b)** the percentage of share of collaboration types. Note that collaboration types indicate edges in the co-authorship network, classified into female-female (FF), male-male (MM), and female-male (FM) collaborations.

## 3.2 Uni-Author Level Disciplinary Diversity

As an initial step in investigating gender-specific patterns of research background diversity in collaboration, we used normalized Shannon's entropy to examine the extent whereto disciplinary diversity exists in an individual's research background, which allows us to address the first research question (Q1). We excluded authors who contributed to only one discipline and analyzed the disciplinary profiles for a total of 114,313 authors, consisting of 83,880 (73%) men and 30,433 (27%) women. Figure 4 shows the probability density distribution of entropies for researchers' publication history, containing eight main research subfields of AI represented by the topic distribution vector. From the smooth density plot it is observed that all authors, irrespective of their gender, tend to have diverse research backgrounds and almost a balanced publication distribution across different disciplines. However, the number of fields ($n_d$) that each author contributes to could negatively affect their ability to equally contribute to all those fields.

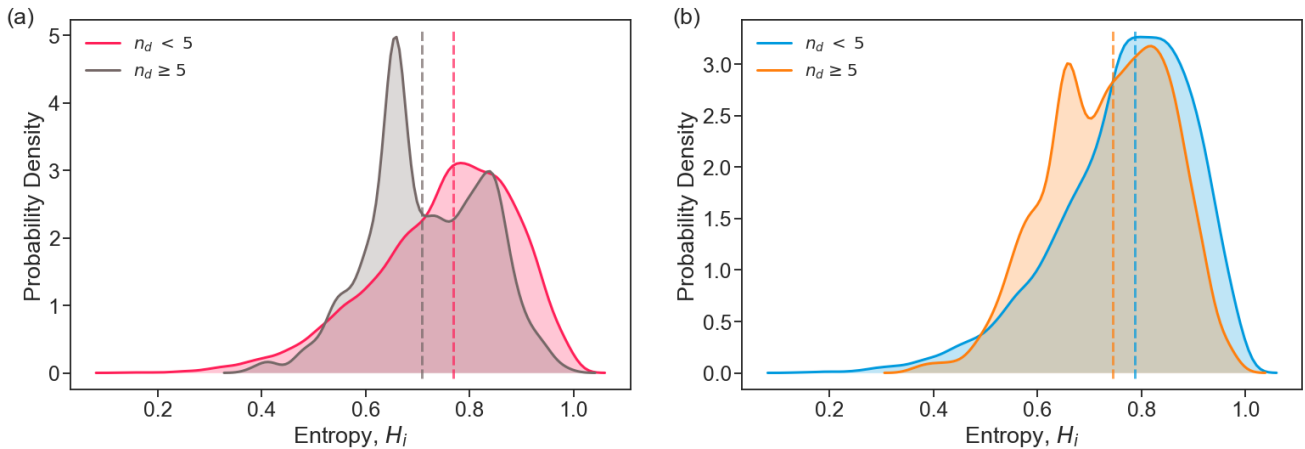

**Figure 4.** Probability density of entropies for **a)** female, and **b)** male researchers. Dashed lines indicate the median of each distribution, and $n_d$ is the number of unique disciplines in which researchers publish. Note that after examining several thresholds for $n_d$, we found 5 as the optimal cutoff.

As depicted in Figure 4-a and -b, both female and male scientists who publish in more fields ($n_d > 5$) have lower entropy values, meaning that they cannot have equal contributions to all the disciplines they are involved in. Within the group of researchers active in more scientific fields ($n_d > 5$), the distribution for both genders is bimodal. While the bimodality for the male group is relatively moderate, the gap between the two peaks for the female group is more pronounced, indicating that the female group is neatly divided into two significant clusters. Notwithstanding the mentioned similarities between the two gender groups, male researchers, on average, have slightly more balanced and diverse research profiles (higher entropy values) than their female counterparts. To statistically support this hypothesis, the one-sided Mann-Whitney U test results revealed that female researchers' median entropy is significantly lower than that of males regardless of the number of fields ($n_d > 5$: Median$_{(female)}$=0.769, Median$_{(male)}$=0.788, p<0.001); ($n_d > 5$: Median$_{(female)}$=0.707, Median$_{(male)}$=0.745, p<0.001). We furthermore performed a permutation test with 10,000 permutations to determine the significance level. This test does not require any distributional and independence assumptions (Edgington, 1980). The results confirmed a significant difference between the medians of the two gender groups.

## 3.3 Bi-Author Level Disciplinary Diversity

As explained in Section 2.2.2., in order to examine whether authors of each gender prefer to collaborate with researchers with similar research profiles, we calculated cosine similarity between researchers' topic distribution vectors for a total of 1,998,318 edges in the co-authorship network, consisting of 1,137,245 male-male (MM), 145,339 female-female (FF), and 715,734 female-male (FM) collaborations. Assuming the authors' topic distribution vectors as individual disciplinary profiles, we studied the dyadic interdisciplinarity between female and male researchers. Figure 5 demonstrates the probability densities of the cosine similarity values ($S_{ij}$) for mixed-gender and same-gender dyads. As seen, cosine similarity values for FF edges are significantly higher than those of MM and FM edges, and mixed-gender dyads tend to have more similar research experience than male dyads. These observations highlight that females have a higher propensity than males to coauthor with other researchers who have similar interdisciplinary research backgrounds. Pairwise comparisons using the Mann-Whitney U test (two-sided) indicated that differences between all groups are statistically significant: (for all dyads including FF and MM, FF and FM, FM and MM: Median=1, p<0.001). However, since the cosine similarity values had a left-skewed distribution and all groups had the same median, we used a one-tailed t-test to determine if the average $S_{ij}$ of FF edges is higher than those of MM and FM edges. Results indicated that differences between all groups are statistically significant, suggesting higher disciplinary homophily levels among FF and FM dyads (FF and MM: Mean$_{(FF)}$= 0.986, Mean$_{(MM)}$= 0.978, p<0.001); (FF and FM: Mean$_{(FF)}$=0.986, Mean$_{(FM)}$=0.981, p<0.001); (FM and MM: Mean$_{(FM)}$=0.981, Mean$_{(MM)}$=0.978, p<0.001). Similar to section 3.2, a permutation test (10,000 permutations) was performed to identify the significance level. The above analysis and our further investigation of team disciplinary diversity in the next section address our second research question (Q2) concerning the existence of disciplinary homophily in scientific collaborations.

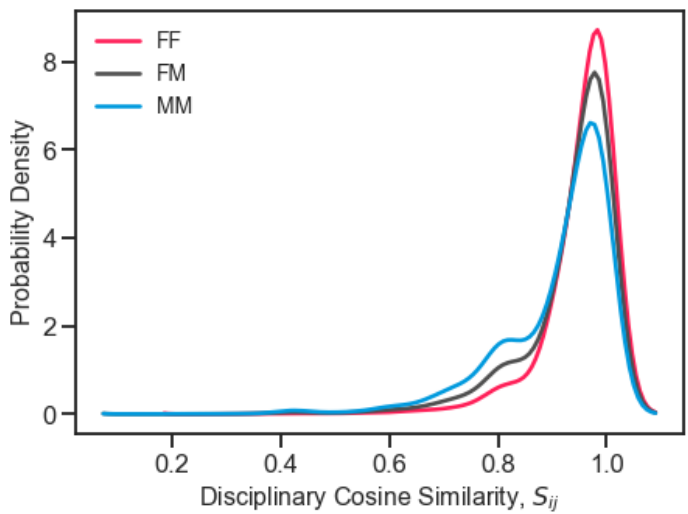

**Figure 5.** Probability density of cosine similarity ($S_{ij}$) for female-female (FF), female-male (FM), and male-male (MM) collaboration types.

## 3.4 Team-Level Disciplinary Diversity

In order to study team disciplinary diversity based on team members' primary disciplines, we applied the following strategy: (1) we considered the total number of unique team size ($p$) as the population and determined an optimal sample size for each population, assuming 95% confidence interval and 1% margin of error, (2) we generated randomized teams by applying random sampling without replacement from all authors in the real dataset, (3) we calculated within-group entropy ($\widetilde{H}_P$) for both real and randomized teams, (4) for each unique team size ($|p|$) in the randomized data, we calculated the average ($\mu_{|p|}^{(H)}$) and standard deviation ($\sigma_{|p|}^{(H)}$) of the entropy values, (5) we compared the real teams with randomized teams by calculating the z-score ($z_p^{(H)}$), which is defined as:

$$z_p^{(H)} = \frac{\widetilde{H}_p - \mu_{|p|}^{(H)}}{\sigma_{|p|}^{(H)}}, \quad (6)$$

and, (6) similarly, we replicated our analysis while using $\widetilde{S}_p$ instead of $\widetilde{H}_P$ and the z-score ($z_p^{(S)}$) to investigate team diversity based on team members' disciplinary profiles. Thereby,

$$z_p^{(S)} = \frac{\widetilde{S}_p - \mu_{|p|}^{(S)}}{\sigma_{|p|}^{(S)}} \quad (7)$$

Figure 6 shows the probability densities of $z_p^{(H)}$ and $z_p^{(S)}$ for female-only, male-only, and mixed-gender teams. In the figure, z-scores indicate how many standard deviations the values of $\widetilde{H}_P$ and $\widetilde{S}_p$ diverge from what is expected by random. It is observed that for all collaboration teams, the distribution of $z_p^{(H)}$ is concentrated on negative values, suggesting that $\widetilde{H}_P$ values for most of the teams are lower than what is expected for randomized groups. On the other hand, the distribution of $z_p^{(S)}$ is concentrated on positive values, meaning that teams with high $\widetilde{S}_p$ are more common in the actual dataset than randomized. Additionally, the probability density of both $z_p^{(H)}$ and $z_p^{(S)}$ for female-only teams tend to be higher compared to the other groups. These observations highlight that although co-authorship teams, in general, tend to be composed of researchers with more similar disciplinary profiles, this preference is stronger among females, who have more homophilous collaborations compared to their male counterparts.

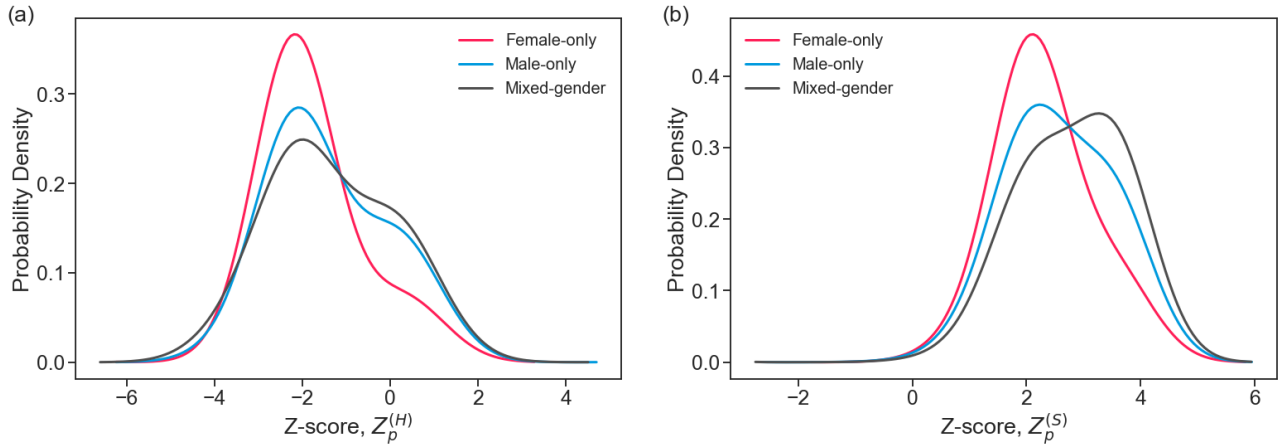

**Figure 6. a)** Probability densities of $z_p^{(H)}$ for female-only, male-only, and mixed-gender teams, and **b)** probability densities of $z_p^{(S)}$ for female-only, male-only, and mixed-gender teams. The z-scores show the deviations from expected values of $\widetilde{H}_P$ and $\widetilde{S}_p$ in the randomized teams.

## 3.5 Scientific Performance and Collaboration Diversity Relationship

In this section, the last research question (Q3) is addressed by analyzing the relationship between female and male researchers' scientific performance and their collaboration diversity. First, using core-periphery decomposition, all network nodes, i.e. researchers, were divided into core and periphery nodes (see section 2.2.2. for details). Out of the 114,371 authors, 5,720 researchers (5%) were associated as core researchers, including 4,197 males and 1,523 females. It was observed that male scientists have more brokerage roles in the network as their average betweenness centrality is 1.8 times higher than women's. However, the proportion of core females has increased by over 23% from 2000 to 2019, indicating that women are occupying more influential positions in tandem with men in the AI research community.

Next, we studied the impact of having core roles within the co-authorship network on research performance and experience. Figure 7 shows the distribution of average citation counts, as a proxy for research impact, for female and male core/periphery researchers. As observed, the core researchers' distribution moves above the periphery researchers' distribution, regardless of the gender, suggesting a positive relationship between occupying core positions in the co-authorship network and research impact in terms of citation counts. Notably, from Figure 7-a and -b, it is seen that the curves for core and periphery researchers are almost the same for female and male authors.

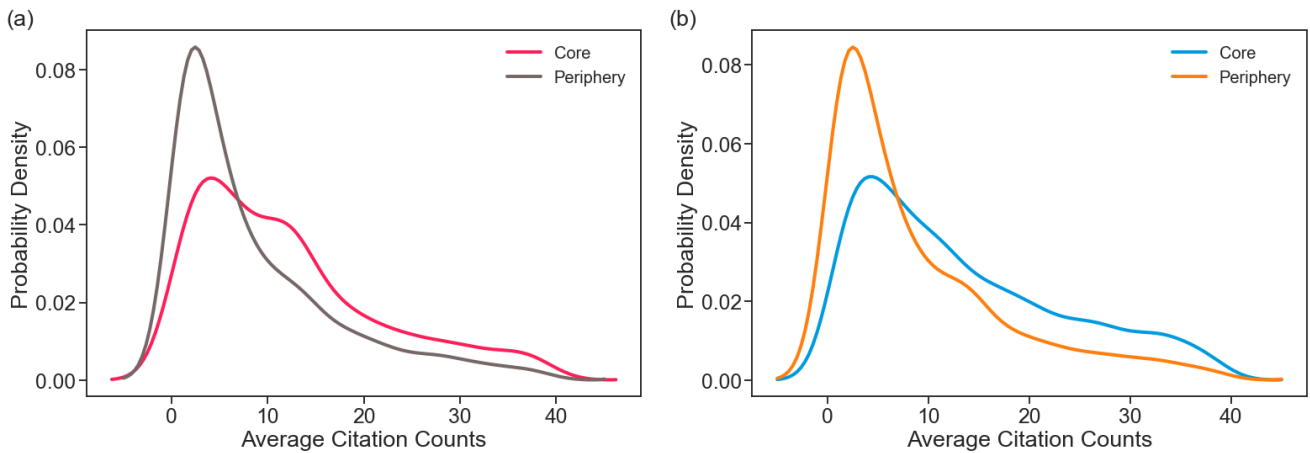

**Figure 7.** Average citation counts distribution for **a)** female, and **b)** male researchers.

We also investigated the relationship between research output, measured by the number of publications, and researchers' brokerage role in the co-authorship network, measured by betweenness centrality, for female and male researchers (Figure 8). The same patterns were observed here as well, indicating a positive relationship between having core positions and research output for both female and male researchers. Additionally, the core/periphery density curves were comparable for female and male researchers (Figure 8-a and -b).

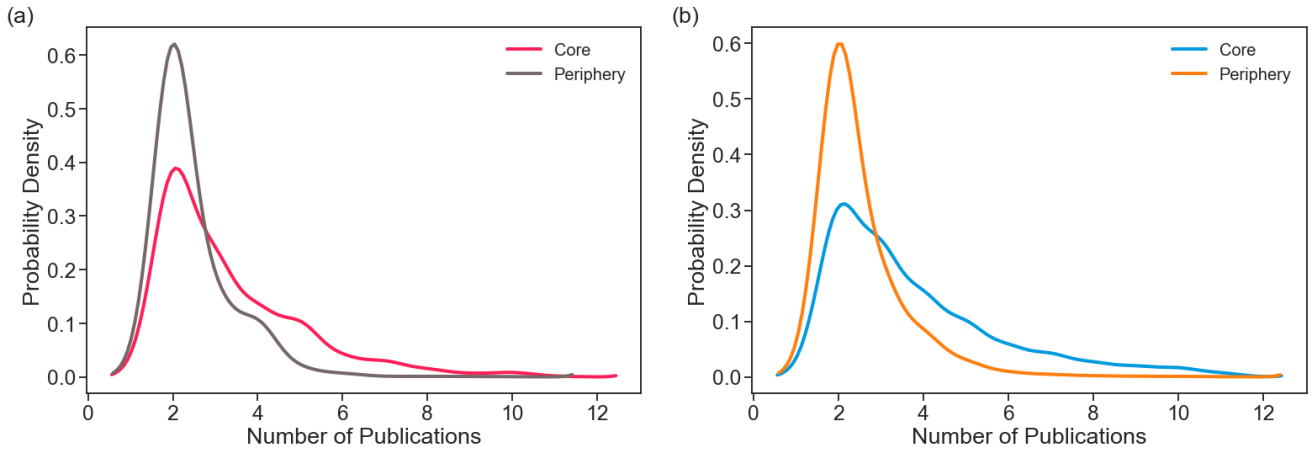

**Figure 8.** Number of publications distribution for **a)** female, and **b)** male researchers.

Figure 9 shows the probability density distributions of researchers' career length as a proxy for academic experience, measured by the difference between authors' first and last publications' dates in the dataset. As seen, the distributions of the core researchers' academic experience for both female and male researchers place above those of the periphery researchers. This suggests a positive relationship between having a core brokerage position in the co-authorship network and longer academic experience, regardless of gender. We validated our findings by performing the one-sided Mann-Whitney U test which resulted in significant differences in all cases ($p < 0.001$).

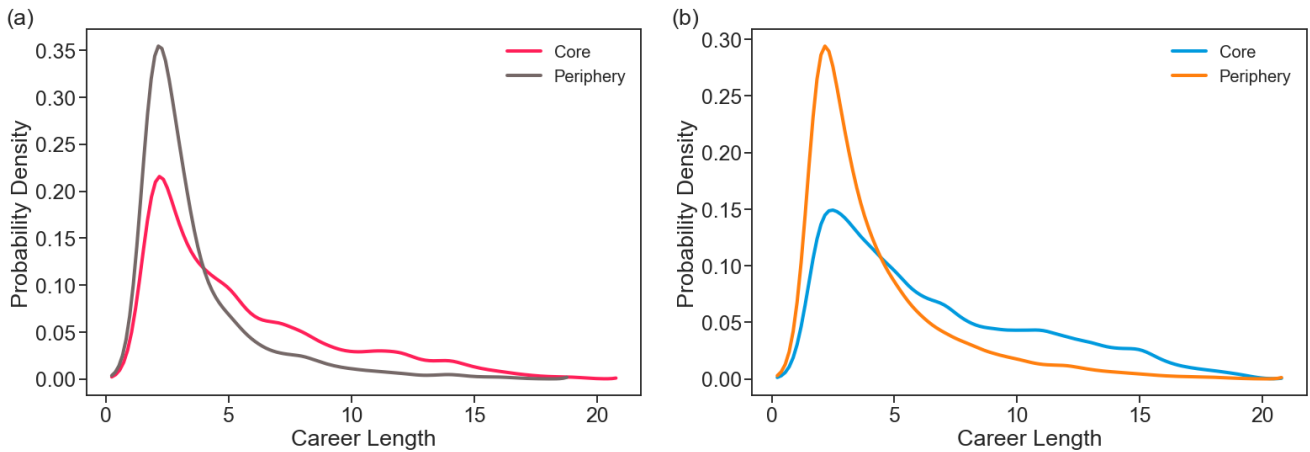

**Figure 9.** Career length distribution for **a)** female, and **b)** male researchers.

Motivated by these findings, we did further statistical tests to examine whether there are differences in research performance and experience between female and male researchers who are holding core positions in the network. The two-sided Mann-Whitney U test revealed that the difference between men and women in core positions was statistically significant in all cases ($p < 0.001$). Furthermore, the one-tailed t-test was performed to compare research performance between the two groups. Pairwise comparisons between the two groups showed that means of the career length and average citation counts of core females were significantly lower than those of core males, and that women in core positions, on average, publish less than their male counterparts (Average raw citation counts: $Mean_{(core\ female)}=20.584$, $Mean_{(core\ male)}=28.236$, $p<0.001$); (Career length: $Mean_{(core\ female)}=4.912$, $Mean_{(core\ male)}=6.093$, $p<0.001$); (Number of publications:

Mean$_{(core\ female)}$=3.462, Mean$_{(core\ male)}$=4.126, p<0.001). We moreover calculated field normalized citation counts and observed similar relationships as the ones seen for the average citation count. Our findings show that authors who collaborate with diverse groups are more likely to have higher seniority levels and scientific performance in terms of both quantity and impact. Weaker research performance was observed for core females compared to the core male researchers; however, it is worthy to mention that having shorter career length as well as having a fewer number of female researchers in the AI scientific ecosystem, as suggested by our data, may have a negative impact on the academic performance of female researchers.

Finally, we closely investigated the collaborative relationships among top core AI researchers, defined as the top-0.1% researchers with the highest betweenness centrality (N=114, 15 females and 99 males). Figure 10-a shows their co-authorship network. The red nodes represent female researchers and the blue nodes represent male researchers. The higher the betweenness centrality value, the bigger the node is in the network. Edges show co-authorship relationships, i.e. if two authors have a joint paper they are directly connected in the graph, and the thickness of edges indicates the collaboration frequency. As seen, not only is the network male dominant, but the top core researchers are also male, i.e. the biggest nodes in the graph. In addition, the gender homophily effect can be observed among both genders meaning that they seem to be collaborating more with researchers of the same gender as theirs. Nevertheless, some female core researchers are creating mixed-gender clusters by joining to male-dominant clusters. Figure 10-b compares the performance of top core AI researchers. As one may observe, top male researchers have more direct connection than their female counterparts, measured by the average degree, however, the gap is negligible. Nonetheless, in terms of the average number of publications, the male researchers produced almost twice the female researchers. Moreover, papers published by males have been more cited, according to their average h-index and i-10 index. These observations are akin to our findings from the core-periphery analysis, indicating the weaker academic performance for women in brokerage roles in the AI community. Regardless of gender, these top core researchers are mostly active in health and biology, partially explaining their high number of direct connections/co-authors. We also compared the ratio of their AI publications in the Scopus database over their total number of publications. The average ratio is slightly higher for men, 4.1% vs 3.7% with 95% confidence intervals of [2.9%, 5.3%] and [2.5%, 5.1%] for males and females respectively.

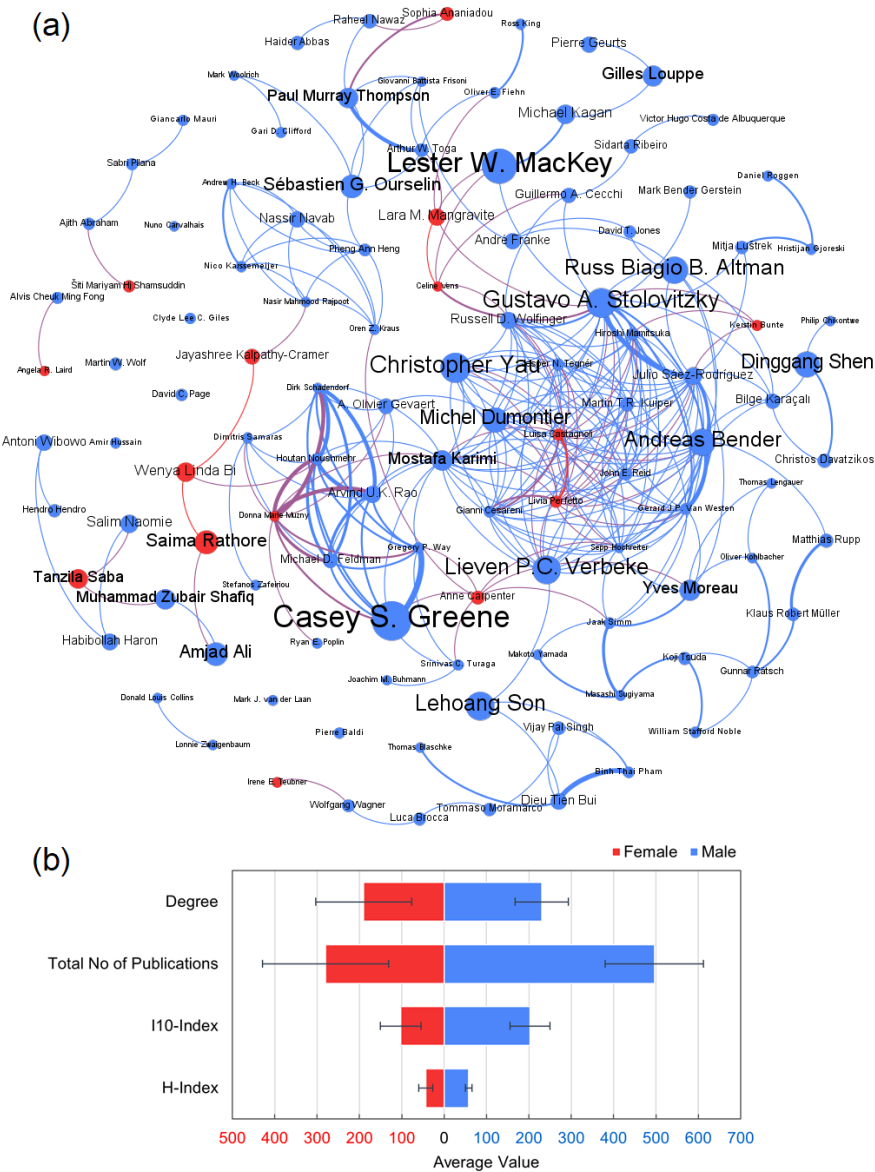

**Figure 10. a)** The co-authorship network of most central AI researchers, i.e. the top-0.1% researchers with the highest betweenness centrality (N=114, 15 females and 99 males). Female nodes are colored red and male nodes blue, and **b)** comparison of male and female AI researchers based on their degree, total number of publications, h-index, and i10-index. The error bar represents a 95% confidence interval for the mean value per gender.

## 4. Discussion

Despite notable progress, gender disparity in science and scientific activities remains, calling for a more systematic and comprehensive approach to tackle and investigate the problem. Many initiatives have endeavored over the last decades to fill the gender gap in the academic and research community. As a highly interdisciplinary and evolving domain, artificial intelligence is constantly affecting many angles of human life, not only attracting the attention of researchers but also the governments and decision-/policymakers. This study complements existing literature on gender differences in scientific collaboration by analyzing disciplinary diversity among female and male researchers at the uni-author, bi-author, and team levels, as well as assessing the influence of disciplinary homophily or diversity on scientific research team formation. We applied multiple techniques, i.e. natural language processing, machine learning, social network analysis, and statistical analysis, to comprehensively investigate gender-specific patterns in the AI scientific ecosystem between 2000 and 2019. To the best of the authors' knowledge, this is the first study that addresses gender differences in the propensity to collaborate concerning the similarity or dissimilarity of researchers' academic backgrounds in a highly-interdisciplinary and evolving field such as AI.

In line with general conclusions of a recent report on the existence of a global gender gap in AI published by the World Economic Forum (World Economic Forum, 2018), we also found that the number of male researchers publishing in the AI domain has been constantly higher than their female counterparts in the entire examined period, with an increasing trend observed for both genders. Notwithstanding, our findings suggest that male and female AI researchers are collaborating together more, indicated by the increasing trend for mixed-gender collaborations. This property was recently observed in some other interdisciplinary fields such as nanotechnology (Ghiasi et al., 2018). This property, along with our observed higher preference of female researchers toward collaboration, may facilitate the knowledge transfer between genders, resulting in generating more female AI superstar researchers that itself could inspire and engage more women at the forefront of the AI domain. Thus, our finding provides new insights into the changing role of female AI scientists and their active involvement in the field, which is in line with the high-level objectives of the frontline countries investing in AI and partially confirms the effectiveness of the policies taken to generate and support women AI leaders. As an example, in Canada, a new initiative has recently been put in place to tackle diversity in AI and foster the next generation of female leaders (CIFAR, 2019), as only 14% of Canadian AI researchers are women (Shelementaicom, 2019). In our dataset, we observed a 22% share for female Canadian AI researchers.

Analyzing male and female scientists' disciplinary diversity revealed that although female scientists have slightly less balanced and less diverse research profiles than males, all authors, regardless of gender, tend to have a wide range of experience in different AI subject areas. In contrast, the tendency to have equal contribution to all fields is decreased as they engage in more research areas. This is in accordance with Feng and Kirkley (2020), demonstrating that individual-level disciplinary diversity is a common indicator of interdisciplinary research collaborations. Generally, several studies denoted researchers prefer working with others who have a certain degree of similarity between different characteristics (AlShebli et al., 2018; Holman & Morandin, 2019). We also found that scientific collaborations within the AI community are more likely to occur between authors from similar research backgrounds, and this implies that uni-author level disciplinary diversity, i.e. individual researchers with several skill sets from various disciplines, plays a significant role in the formation of interdisciplinary collaborations. To borrow from Bunderson & Sutcliffe (2002), teams composed of interdisciplinary individuals might be better capable of sharing their knowledge and overcoming common collaboration barriers, e.g. different mindsets, lack of mutual understanding, and communication problems, which could lead to higher team effectiveness. These findings provide insights for both researchers and organizations about team composition characteristics in the AI scientific ecosystem from a gender perspective.

Our analysis of gender-specific collaborative patterns confirmed the existence of disciplinary homophily at both bi-author and team levels, consistent with previous findings by Feng and Kirkley (2020). Broadly, AI researchers are inclined to collaborate with authors who have similar disciplinary profiles; therefore, our results complement previous studies reporting the presence of homophily in academic collaborations in terms of gender (Holman & Morandin, 2019; Jadidi et al., 2018), affiliation, discipline, and ethnicity (AlShebli et al., 2018). Further, we found that women exhibit a stronger tendency toward disciplinary homophily compared to men in the AI scientific ecosystem. Female scientists tend to form more homophilous collaborative ties, i.e. interacting more with scientists from similar research backgrounds, and this tendency becomes stronger when it comes to female-female collaborations. One plausible explanation for this homophily effect could be the existence of a gender bias in terms of collaboration with like-minded people that could facilitate information/knowledge transmission and improve team productivity. However, we assume that researchers who are primarily collaborating with those of the same disciplinary profile, in a highly evolving and interdisciplinary ecosystem such as AI, might be less exposed to the cutting-edge advancements in the field that could affect the gender gap in time. Therefore, it is suggested that researchers of both genders tend to consider diverse types of scientific collaboration at all levels (AlShebli et al., 2018).

Lastly, the core-periphery analysis indicated a significant positive association between having diverse collaboration and scientific performance and experience. In line with the existing literature (Abbasi et al., 2012; Feng & Kirkley, 2020; Yan & Ding, 2009), our findings demonstrated that having core positions within the scientific network could positively enhance scholars' academic performance and, as a result, their success. Core researchers can play an exceedingly important role in developing new collaboration ties with researchers from diverse communities, enabling them to reap the benefits of a wider range of knowledge and experience as well as higher scientific performance. In addition, our findings suggest tangible differences between the two genders occupying core network positions. On average, core female scientists have lower seniority levels and scientific performance in terms of both productivity and impact. We speculate that this lower performance might be due to the fact that women are more likely to leave research careers or pause their scientific activities for a plethora of reasons such as family obligations (Ramos et al., 2015), inadequate female role models (Nelson & Rogers, 2003), masculinist working atmosphere, and professional preferences (M. T. Wang & Degol, 2017). If true, this would aid in explaining fewer influential women in the AI community, impeding females' academic success.

## 5. Conclusion

Undoubtedly being the main driver of many emerging technologies, artificial intelligence is changing the world and many aspects of the way we live, impacting tremendously our future. As an evolving and complex ecosystem that is attracting significant investments annually, training leading AI researchers is also complex and costly. In this study, we comprehensively analyzed gender-specific patterns in the AI scientific ecosystem from 2000 to 2019 and shed light on gender differences in collaborative behavior and how these differences could impact academic success. We also highlighted that disciplinary homophily is one of the main team composition characteristics in the AI scientific ecosystem. In spite of the gap observed between the female and male researchers active in the field, we found promising indications of a general will towards filling the gender gap in the AI ecosystem. The differences observed between genders in performing scientific activities and collaboration are of crucial importance. We believe our analysis provides a comprehensive picture of the gender-specific patterns in AI that could help the policymakers to adjust existing or set new strategies to support female researchers more sustainably and efficiently by considering all the aspects of the differences between females and males not only in initiating, leading, and maintaining collaboration and scientific activities but other influencing factors such as social and family responsibilities that could impact their professional career.

## 6. Limitations and Future Work

We focused on gender-specific patterns in the AI field, as a highly interdisciplinary male-dominated field (World Economic Forum, 2018). Future research could analyze other emerging technologies to compare the results. We applied an automatic gender assignment algorithm and excluded authors for which we were unable to assign gender. Future research could explore other gender assignment strategies such as manual and/or semi-automatic. In this study, we analyzed collaboration patterns among researchers through a co-authorship network. Although the co-authorship network is regarded as one of the most common and tangible indicators to measure scientific collaboration, it cannot capture all collaboration types inasmuch as collaborative efforts do not always result in a joint publication (Katz & Martin, 1997). Another future direction could be addressing this issue by taking other approaches to measure research collaboration. We used betweenness centrality to measure researchers' brokerage role in the co-authorship network. Future studies may consider other centrality measures and analyze their relationships with researchers' scientific performance. Additionally, we considered the time difference between the authors' first and last publications as a proxy for their career age as we did not have data on researchers' age and/or seniority level. One may use other proxies for authors' seniority level in future research and compare the results. Lastly, we used articles' abstracts and titles to infer researchers' disciplinary profiles. Future research could further explore the authors' research disciplines by taking the entire publications' text into the account.

### Data Availability

Data is available upon request. Please contact the corresponding author.

### Author Contributions

Conceptualization: AE. Data collection: AH. Data curation: AH. Methodology and design of experiments: AH, AE. Experimentation and data analysis: AH. Interpretation of results: AH, AS, AE. Resources: AS. Supervision: AE, AS. Wrote the manuscript: AH, AS, AE.

### Competing Interests

The authors have no competing interests.